\documentclass[a4paper]{article}

\usepackage{graphicx}
\usepackage{amssymb}
\usepackage{amsmath}
\usepackage{subfigure}
\usepackage{fancyhdr}

\usepackage[left=3cm,right=3cm,top=1.5cm,foot=1cm]{geometry}

\usepackage{hyperref}
\hypersetup{
    unicode=true,
        pdfborder={0 0 0},
        pdftitle={
            Efficient multicast data transfer with congestion control using
                dynamic source channels
        },
        pdfsubject={
            Efficient multicast data transfer with congestion control using
                dynamic source channels
        },
        pdfauthor={
            Vincent Lucas,
            Jean-Jacques Pansiot,
            Dominique Grad and
            Benoit Hilt
        },
        pdfcreator={Vincent Lucas},
        pdfproducer={Vincent Lucas},
        pdfkeywords={multicast, congestion control, FEC}
}

\begin{document}
    \begin{center}
        {\large
            $LSIIT$ - Universit\'e de Strasbourg / $CNRS$
            \\
            \vspace{4cm}
            {\huge
                Efficient multicast data transfer\\
                    with congestion control\\
                    using dynamic source channels
            }
            \\
            \vspace{2cm}
            {\Large
                Vincent Lucas$^1$,
                Jean-Jacques Pansiot$^1$,
                Dominique Grad$^1$
                and Beno\^it Hilt$^2$
            }
            \\
            \vspace{1cm}
            ($^1$) LSIIT - Universit\'e de Strasbourg - CNRS, France\\
            \textit{\{lucas, pansiot, dominique.grad\}@unistra.fr}\\
            ($^2$) Laboratoire MIPS - Universit\'e de Haute Alsace\\
            \textit{benoit.hilt@uha.fr}
            \\
            \vspace{5cm}
            {\Large
                Research report - $23^{rd}$ april 2010
            }
        }
    \end{center}
    \newpage

\title{
    Efficient multicast data transfer with congestion control using dynamic
        source channels
}

\author{%
    Vincent Lucas$^1$,
    Jean-Jacques Pansiot$^1$,
    Dominique Grad$^1$
    and Beno\^it Hilt$^2$
    \\
    \\
    ($^1$) LSIIT - Universit\'e de Strasbourg - CNRS, France\\
    \textit{\{lucas, pansiot, dominique.grad\}@unistra.fr}\\
    ($^2$) Laboratoire MIPS - Universit\'e de Haute Alsace\\
    \textit{benoit.hilt@uha.fr}
}

\maketitle

%    - I usually write the abstract last
%    - Used by program committee members to decide which papers to read
%    - Four sentences [Kent Beck]
%    -- 1. State the problem
%    -- 2. Say why it’s an interesting problem
%    -- 3. Say what your solution achieves
%    -- 4. Say what follows from your solution
\begin{abstract}
    The most efficient receiver-driven multicast congestion control protocols
    use dynamic channels. This means that each group has a cyclic rate variation
    with a continuously decreasing phase.  Despite promising results in terms of
    fairness, using efficiently these dynamic groups could be a challenging task
    for application programmers.  This paper presents a sequencer which maps out
    application data to dynamic groups in an optimal way.  Multiple applications
    such as file transfer or video streaming, can use this sequencer, thanks to
    a simple $API$ usable with any buffer containing the most important data
    first. To evaluate this solution, we designed a file transfer software using
    a $FEC$ encoding.  Results show the sequencer optimal behavior and the file
    transfer efficiency, as a single download generates only little more
    overhead than $TCP$.  Moreover, download time is almost independent of the
    number of receivers, and is already faster than $TCP$ with $2$ competing
    downloads.
\end{abstract}

%\begin{keywords}
%    sequencer, scheduler, $FEC$, multicast, file transfer
%\end{keywords}

%        1. Describe the problem
%            -  Use an example to introduce the problem
%        2.  State your contributions
%            - Write the list of contributions first
%            - The list of contributions drives the entire paper: the paper
%                substantiates the claims you have made
%            - Reader thinks “gosh, if they can really deliver this, that’s be
%                exciting; I’d better read on”
%        ...and that is all
%        ONE PAGE!
\section{Introduction}
    Most multicast flows such as IPTV are streamed at a constant rate using
    $UDP$.  To fairly share the bandwidth with $TCP$ streams, several multicast
    congestion controls have been proposed. Highly scalable ones are layered and
    receiver-driven, thus each receiver can adapt its reception rate to the
    network capacity.  In the first propositions, such as
    $RLM$~\cite{mccanne_receiver-driven_1996},
    $RLC$~\cite{vicisano_tcp-like_1998} or
    \textit{FLID-SL}~\cite{byers_flid-dl:_2002} , the layers correspond to
    multicast groups sent at a constant rate.  Each receiver joins or leaves new
    multicast groups to respectively increase or decrease its reception rate.
    But, to avoid the multicast leave time latency problem, a new approach using
    dynamic rate groups has been proposed, such as
    \textit{FLID-DL}~\cite{byers_flid-dl:_2002}, $WEBRC$~\cite{luby_wave_2002}
    or $M2C$~\cite{lucas_fair_2009}.  Each group begins to send at a high rate
    and slows down progressively until becoming quiescent. Thereby, a receiver
    has only to wait in order to reduce its reception rate. This solves the
    leave latency problem as a receiver only leaves a multicast group when it
    becomes quiescent.  Despite promising results in terms of fairness, using
    efficiently these dynamic groups is a challenging task and as far as we
    know, no solution to this problem has been proposed yet.
    %\\
    This paper presents a sequencer which maps out application data to dynamic
    groups in an optimal way.  Multiple applications such as file transfer or
    video streaming, can use this sequencer, thanks to a simple $API$ usable
    with any buffer containing the most important data first. Indeed, the
    application does not need to know anything of the dynamic layering
    mechanism and the sequencer is independent of the application as it works
    without knowing the data encoding scheme.
    %\\
    To evaluate this solution, we designed a file transfer software using the
    sequencer and a $FEC$ encoding.  Results show the sequencer optimal behavior
    and the file transfer efficiency, as a single download generates barely more
    overhead than $TCP$.  Moreover, download time is almost independent of the
    number of receivers, and is already faster than $TCP$ with $2$ competing
    downloads.

\section{Addressed issues}
    An efficient massively scalable multicast congestion control uses several
    multicast groups with dynamic layering.  Despite promising results in terms
    of fairness, using efficiently these dynamic groups is a challenging task
    and no solution to this problem has been proposed yet.
    %\\
    The main purpose of this paper is to design a simple $API$, which enables to
    use dynamic multicast congestion control in an optimal way, and takes as
    parameter a buffer containing the most important data first. Therefore, a
    receiver which receives only $N$\% of the source rate, obtains the first
    $N\%$ of the application buffer.  Moreover, various applications can use
    this $API$ without knowing anything about the dynamic layering used.

\section{Multicast congestion control}
    This section gives the background on multicast congestion control, necessary
    to understand the design of an efficient sequencer.  This description
    is based on the receiver-driven \textit{Multicast Congestion Control}
    ($M2C$)~\cite{lucas_fair_2009} protocol, but the sequencer is also
    compatible with \textit{FLID-DL}~\cite{byers_flid-dl:_2002} and
    $WEBRC$~\cite{luby_wave_2002}.

    \subsection{The source part}
        The source part is quite similar to $WEBRC$ and sends data using dynamic
        multicast groups (cf.  figure~\ref{fig:mcc_recv_rate}). Each group
        begins to send at a high rate and then continuously decreases its rate.
        Once a group reaches a low rate threshold, it stops sending data and
        becomes quiescent.  Therefore, a receiver just waits to decrease its
        reception rate and leaves only quiescent groups.  Only the base group,
        which cumulative rate%
        \footnote{
            The cumulative rate for the group of level $L$ corresponds to
                $\sum_{l=0}^{L}\left(rate_l\right)$.
        }
        is $g0$ in figure~\ref{fig:mcc_recv_rate}, never becomes quiescent.
        %\\
        The source uses a time division named \textit{Time Slot Index} ($TSI$),
        which changes each \textit{Time Slot Duration} ($TSD$) seconds. For
        each change of $TSI$, $K$ groups become quiescent and $K$ new groups
        start at the $K$ highest rates. Let us define $sub\_TSI$ the subdivision
        of the $TSI$ corresponding to a duration of $TSD/K$. This notion of
        $sub\_TSI$ will be used by the sequencer.

    \subsection{The receiver part}
\begin{figure}[!t]
            \centering
            \includegraphics[width=.95\textwidth]
                %{img/mcc/mcc_recv_rate.eps}
                {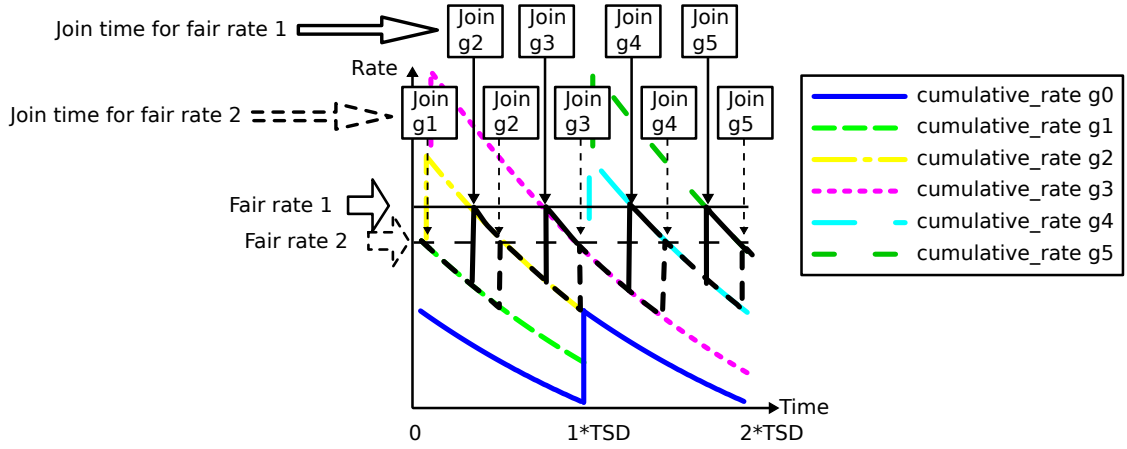}
            \caption{Fine rate granularity due to source dynamic design}
            \label{fig:mcc_recv_rate}
        \end{figure}

        This is the complex part of the multicast congestion control, see
        $M2C$~\cite{lucas_fair_2009} for details.  First, the receiver computes
        the fair rate corresponding to its experienced network conditions. Then,
        the receiver joins multicast groups to receive a rate close to the
        computed fair rate.
        %\\
        Since $M2C$ uses hierarchical cumulative groups, a receiver can join
        group $N$ only after joining all groups from $0$ to $N-1$.  Thereby,
        data sent to the lowest groups in the hierarchy are received by most
        receivers.
        %\\
        Receivers of a static source design can only obtain a set of rates with
        a coarse granularity corresponding to the inter-groups rate granularity.
        Whereas, with dynamic source design a receiver can obtain almost any
        desired rate by joining groups more or less quickly, independently of
        the inter-group rate granularity.  Indeed,
        figure~\ref{fig:mcc_recv_rate} shows that if a receiver can get rate $1$
        by joining group at the shown time, another one can get rate $2$ just by
        delaying its joins, and this with a smaller gap between rates $1$ and
        $2$ than between cumulative rates $g1$ and $g2$.

%        - Explain it as if you were speaking to someone using a whiteboard
%        - Conveying the intuition is primary, not secondary
%        - Once your reader has the intuition, she can follow the details (but
%                not vice versa)
%        - Even if she skips the details, she still takes away something
%           valuable
%        - Do not recapitulate your personal journey of discovery. This route
%           may be soaked with your blood, but that is not interesting to the
%           reader.
%        - Instead, choose the most direct route to the idea.
%         - Introduce the problem, and your idea, using EXAMPLES and only then
%            present the general case
\section{Principle of the sequencer}
    To illustrate the principle of the proposed sequencer, assume that we wish
    to send a video using a dynamic layering congestion control, such as each
    receiver obtains a video resolution corresponding to its reception rate.
    For each video frame, the video software generates a buffer containing the
    different resolutions.  As explained before, the goal is to send the most
    important data to the lowest groups since most of receivers receives it. In
    these conditions, the source application has to map out its data in function
    of the continuously changing rate of each multicast group.
    %\\
    This computation is too specific and too complex for the application level.
    The basic idea presented in this paper proposes to move this mapping
    complexity to a dedicated sequencer easily usable by various applications
    such as file transfer or video streaming.  It provides a simple $API$, which
    only needs a hierarchically encoded buffer, without knowing the buffering
    scheme used by the application. The buffer has just to contain the most
    important bytes at the beginning and the less important at the end.
    %\\
    Then, the sequencer computes the amount of data that each multicast group
    can send, classifies and sequences the application buffer to send the most
    important data at the lowest groups.  Thus, a receiver which receives $N$\%
    of the source rate, obtains the first $N\%$ of the application buffer.

%        - Your introduction makes claims
%        - The body of the paper provides evidence to support each claim
%        - Check each claim in the introduction, identify the evidence, and
%            forward-reference it from the claim
%        - Evidence can be: analysis and comparison, theorems, measurements,
%           case studies
\section{Proposition details}
    This section details how the sequencer splits incoming buffers into
    packets, orders and gives them to the multicast congestion control protocol,
    which schedules them.

    \subsection{Algorithm parameters}
        The sequencer provides a simple $API$ easy to use, which only needs $3$
        parameters:
        \begin{itemize}
            \item{
                The hierarchically encoded $buffer$ to send. It is important to
                    note that the sequencer does not need to know the 
                    encoding scheme used.
            }
            \item{
                The time allowed to send the buffer ($buffer\_time$). For
                    example, if the buffer is a video frame, $buffer\_time$ is
                    the duration of the frame.
                    %\\
                    Concerning file transfer or any other application where the
                    $buffer\_time$ value is not obvious, it can be inferred by
                    the application thanks to the hierarchy used.  Since each
                    receiver must at least receive the first level of the
                    hierarchy, the $buffer\_time$ must correspond to the time
                    needed to send the amount of data of the first level at the
                    minimal congestion control rate%
                    \footnote{
                        The minimal and maximal rates of the congestion control,
                            are fixed when the multicast congestion control
                                session is initiated.
                    }.
                    %\\
                    Also, the $buffer\_time$ is independent of any parameter of
                    the congestion control like $TSD$.
            }
            \item{
                The length of the buffer ($buffer\_length$).  For a video codec,
                    $buffer\_length$ depends on each image compression.  For a
                        file transfer, the ideal $buffer\_length$ is the amount
                        of data sent in $buffer\_time$ at the maximal
                        rate\footnotemark[\value{footnote}].
            }
        \end{itemize}
        The sequencer must query some parameters from the congestion control
        too, such as the amount of data that can be sent for each group during a
        given time slot.
        %\\
        Once all these parameters acquired, the sequencer computes the optimal
        distribution to send the most significant data to the lowest groups
        available.

    \subsection{Packet classifier}
        \begin{figure}
            \centering
            \subfigure[Source dynamic group design]
            {
                \includegraphics[width=0.4\textwidth]
                    %{img/scheduler/algo_scheduler.eps}
                    {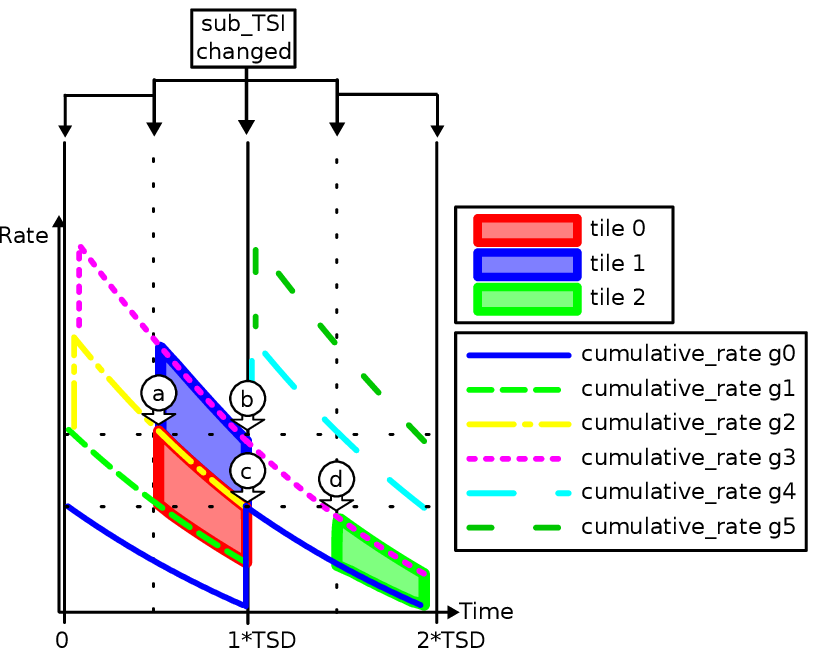}
                \label{fig:algo_scheduler}
            }
            \subfigure[Tile division example]
            {
                \includegraphics[width=0.4\textwidth]
                    %{img/scheduler/advanced_scheduler.eps}
                    {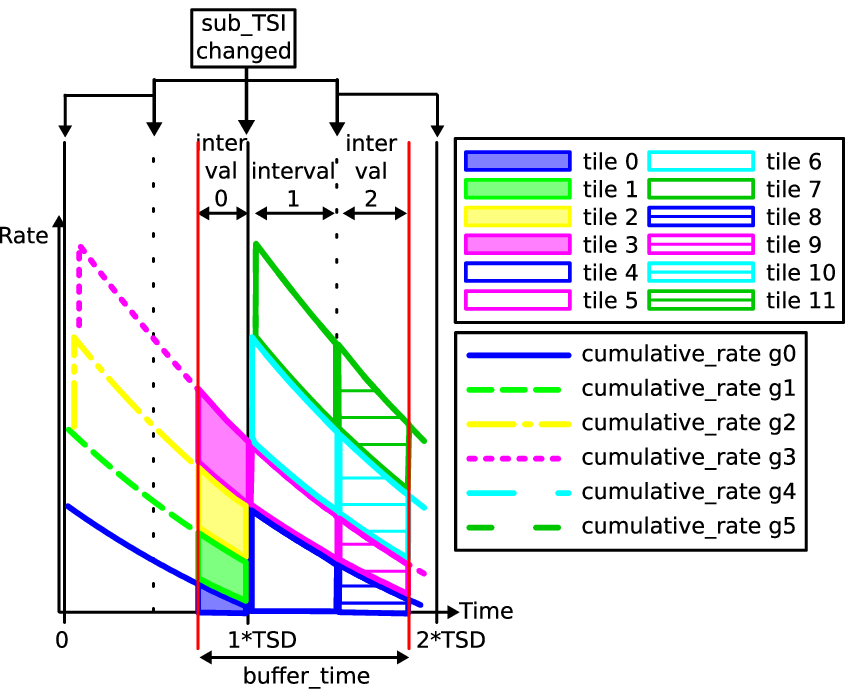}
                \label{fig:advanced_scheduler}
            }
            \subfigure[Algorithm example]
            {
                \includegraphics[width=0.4\textwidth]
                    %{img/scheduler/advanced_scheduler_example.eps}
                    {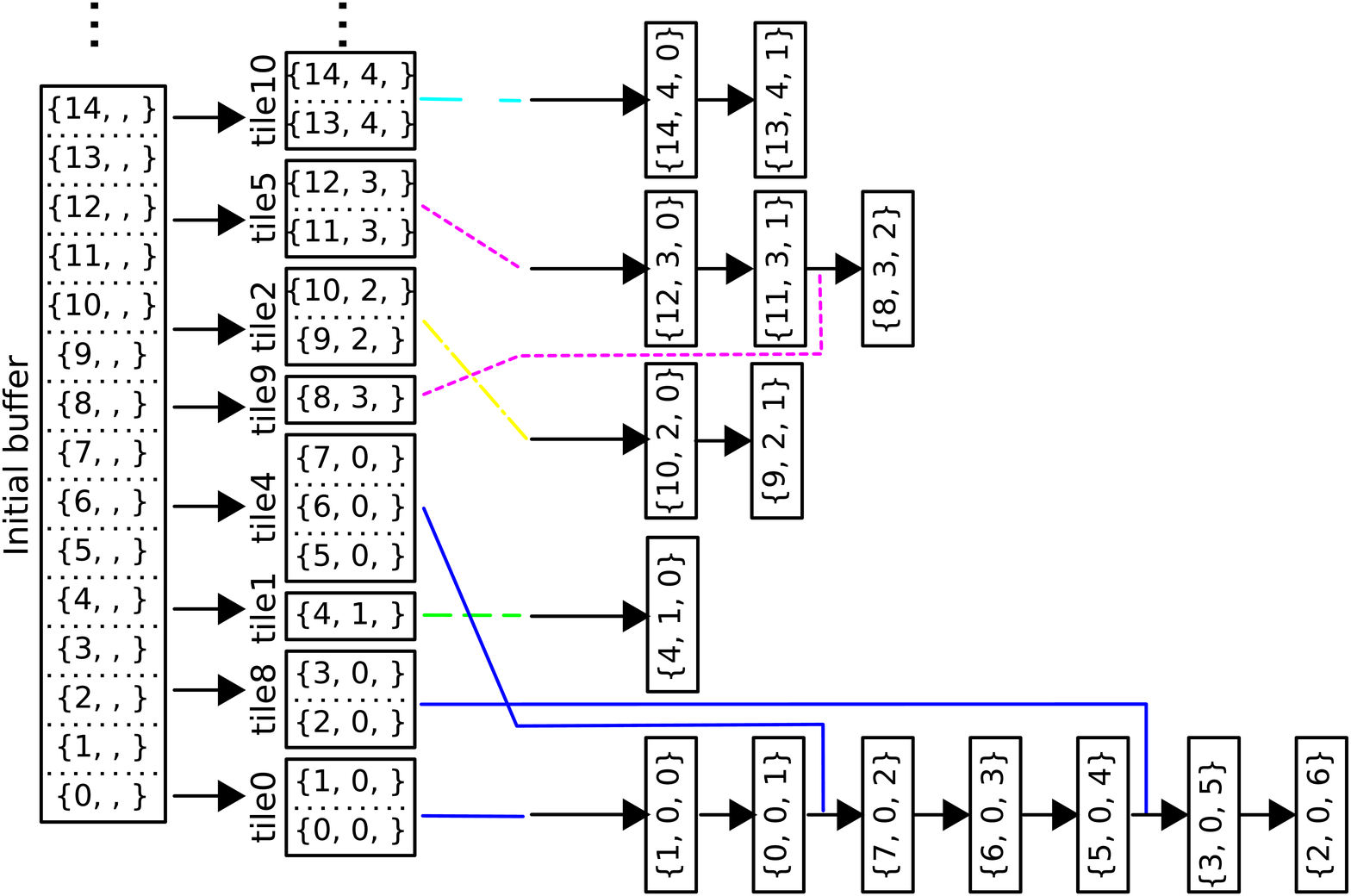}
                \label{fig:advanced_scheduler_example}
            }
            \caption{Sequencer algorithm}
        \end{figure}
        To send the most important data at the lowest group available, the
        sequencer creates a packet hierarchy dependent on time and on the group
        used. The packet hierarchy is subdivided in:
        \begin{itemize}
            \item{
                A coarse grain hierarchy, which splits each group into $tiles$
                    and sorts all tiles in function of their minimal rate. A
                    $tile$ corresponds to a group during a $sub\_TSI$.  It takes
                    $TSD/K$ seconds for the minimal cumulative rate of group
                    $G_N$ to become lower than the maximal cumulative rate of
                    group $G_{N-1}$ during the same period. For example,
                          figure~\ref{fig:algo_scheduler} shows:
                    \begin{itemize}
                        \item{
                            $tile0$ maximal cumulative rate \textcircled{a} is
                                equal to $tile1$ minimal cumulative rate
                                \textcircled{b}. This way, $tile0$ cumulative
                                rate is lower or equal to $tile1$.
                        }
                        \item{
                            $tile0$ minimal cumulative rate \textcircled{c} is
                                equal to $tile2$ maximal cumulative rate
                                \textcircled{d}. This way, $tile0$ cumulative
                                rate is greater or equal to $tile2$.
                        }
                    \end{itemize}
                    The $tile$ order of this example is: $tile 2 < tile 0 <
                        tile 1$. Meaning that the most important data must be
                        send first in $tile2$, then $tile0$, then in $tile1$.
            }
            \item{
                A fine grain hierarchy to sort intra-$tile$ packets.  Depending
                    on the join time, a receiver may get only the end of a
                    $tile$, hence the hierarchy must be sorted in function of
                    decreasing time: i.e. most important packet sent last.
            }
        \end{itemize}

    \subsection{Sequencer algorithm}
        The algorithm goal is to split the buffer into packets and to order
        these packets, sending the most important data at the lowest group
        available (i.e.  figure~\ref{fig:advanced_scheduler_example}).
        \\
        The first thing to do is to split the buffer into \textit{Packet Data
            Units} ($PDU$).  $PDU$ will be identified by the triplet $\left\{j,
            g, s\right\}$ where:
        \begin{itemize}
            \item{
                $j$ is the $PDU$ number%
                    \footnote{
                        In practice, $j$ is an offset relatively to the
                            $buffer$. But for readability purposes, in this
                            paper we use a $PDU$ number corresponding to:
                            \\
                            $PDU\_number := offset / PDU\_length$.
                    }.
                    With $j \in \left[0...J\right[$ and $J$ the number of $PDU$
                        contained in the application buffer. All $PDU$ have
                        the same size, except the last one if the
                        $buffer\_len$ is not a multiple of the default $PDU$
                        size.
            }
            \item{
                $g$ is the group number to which $PDU_j$ will be send. With $g
                    \in \left[0...G\right[$ and $G$ the number of groups.
            }
            \item{
                $s$ is the packet sequence number relative to $g$.  With $s \in
                    \left[0...S_g\right[$ and $S_g$ the number of packets that
                    the group $g$ can send in $buffer\_time$.
            }
        \end{itemize}
        During $buffer\_time$ the same group can be partitioned into several
        tiles.  Indeed, the algorithm splits the $buffer\_time$ into intervals
        defined by $sub\_TSI$, as illustrated by
        figure~\ref{fig:advanced_scheduler}.  Then, it requests the
        corresponding number of packets $S_{g,i}$ that the group $g$ can send
        during interval $i$.  All tiles are sorted by their minimal cumulative
        rates, in order to send most important $PDU$ to the tile having the
        minimal cumulative rate possible. For example in
        figure~\ref{fig:advanced_scheduler_example}:
        \begin{itemize}
            \item{
                Tile 0 will send $PDU_j$ where $j \in \left[0...S_{0,0}\right[$.
            }
            \item{
                Tile 8 will send $PDU_j$ where $j \in
                    \left[S_{0,0}...S_{0,0}+S_{0,1}\right[$.
            }
            \item{
                etc.
            }
        \end{itemize}
        For a given group, the tiles are sent following their chronological
        order: i.e. group $0$ sends first tile $0$, tile $4$ and then tile $8$.
        Finally, the packets of a $tile$ are ordered following the fine grain
        hierarchy.  Inside the same tile, a group continuously decreases its
        send rate, meaning that each tile must send less important $PDU$ first,
        in order to send the most important data at the lowest rate available.

    \subsection{Sequencer header and receiver part}
        The sequencer header is composed of a buffer identifier and an offset
        relative to the application buffer. The receiver stores in a reception
        buffer all the datagrams until one arrives with a new buffer identifier.
        Then, the $API$ provides to the application an access function to
        get a list of all buffer parts received or only the contiguous
        segment starting at offset $0$.

\section{File transfer software design}
    \begin{figure}
        \centering
        \subfigure[Software data flow diagram]
        {%
            \includegraphics[width=0.4\textwidth]
                %{img/soft/software_state_transition.eps}
                {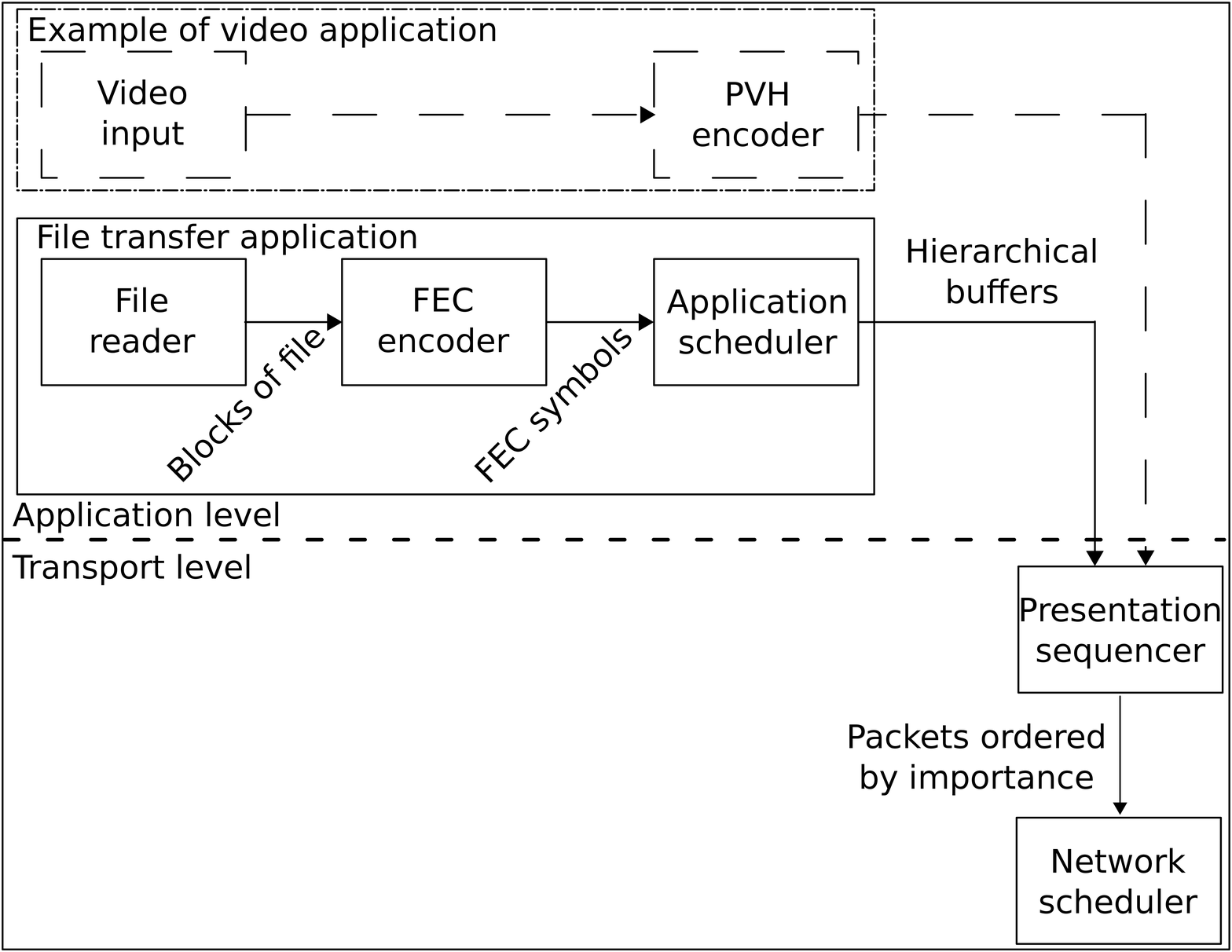}
            \label{fig:soft_data_flow}
        }
        \subfigure[Application scheduler algorithm]
        {%
            \includegraphics[width=0.55\textwidth]
                %{img/soft/application_orderer.eps}
                {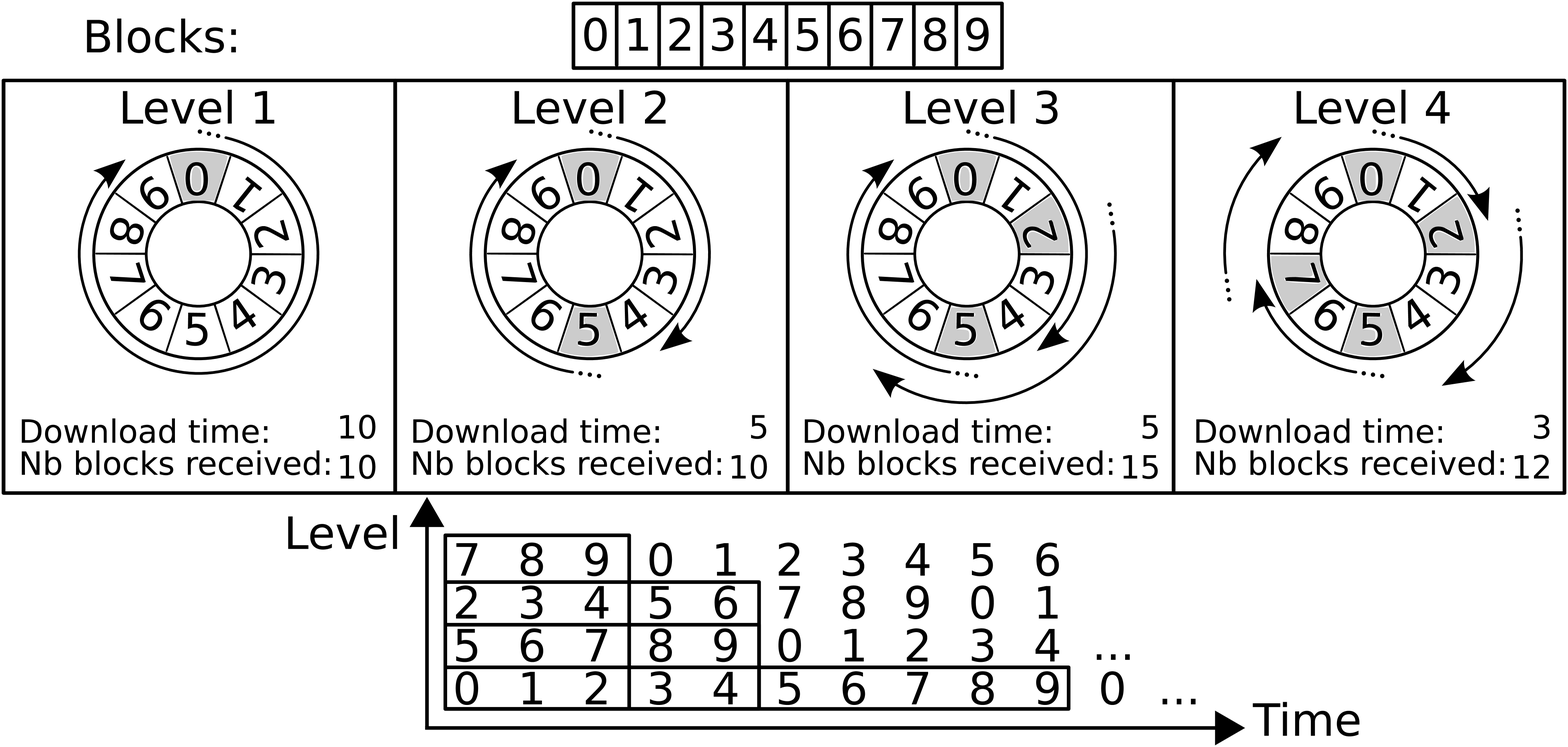}
            \label{fig:application_scheduler}
        }
        \subfigure[Blocks sent before sending a duplicate]
        {%
            \includegraphics[width=0.3\textwidth]
                %{img/soft/duplicated_blocks.eps}
                {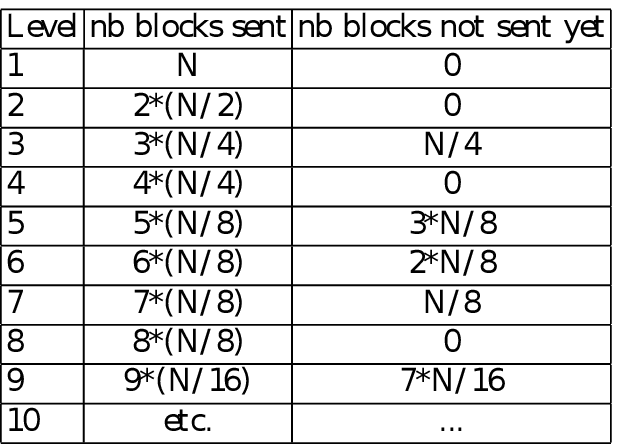}
            \label{table:blocks_sent}
        }
        \subfigure[4000 blocks scheduled]
        {%
                \includegraphics[width=0.3\textwidth]
                    %{img/theorical_res/sched_only/sched_time_nb_blocks_4000.eps}
                    {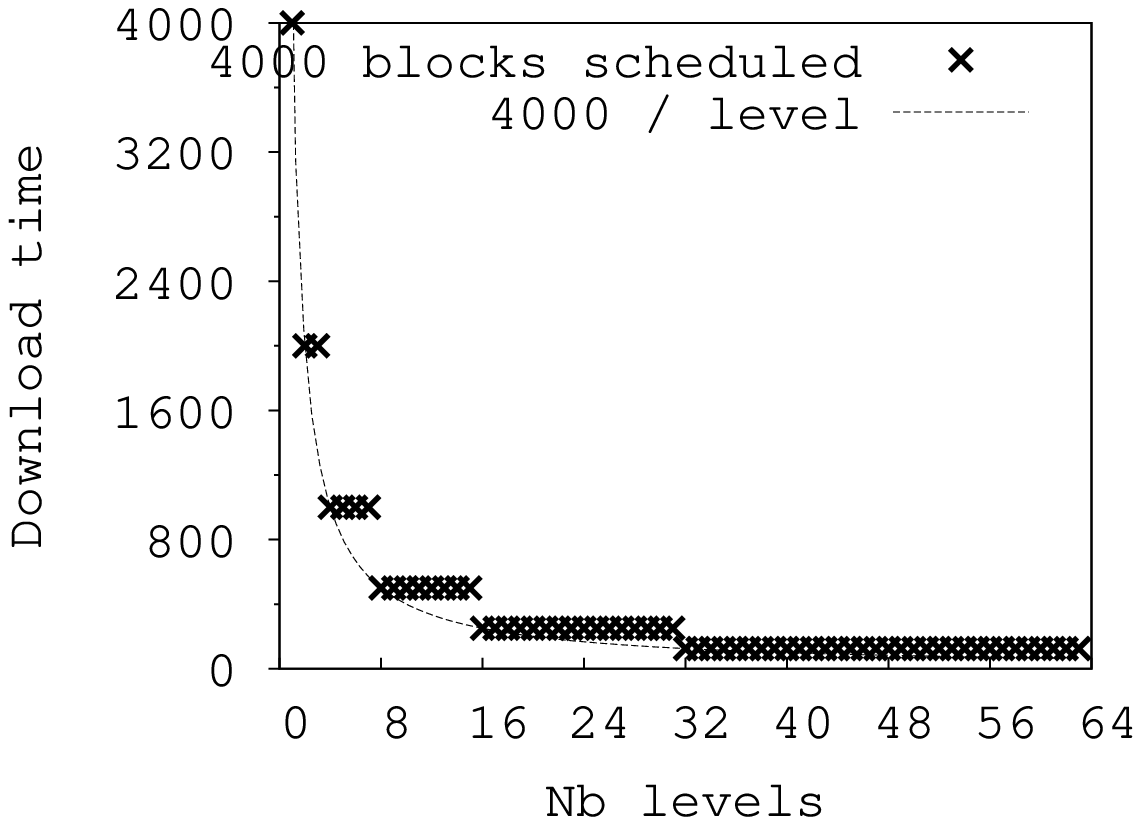}
                \label{fig:theorical_res_sched}
        }
        \subfigure[4000 blocks scheduled with $FEC$ ($k=4000$, $n=8000$,
                $\epsilon=0$)]
        {%
                \includegraphics[width=0.3\textwidth]
                    %{img/theorical_res/sched_div_2/sched_time_nb_blocks_8000.eps}
                    {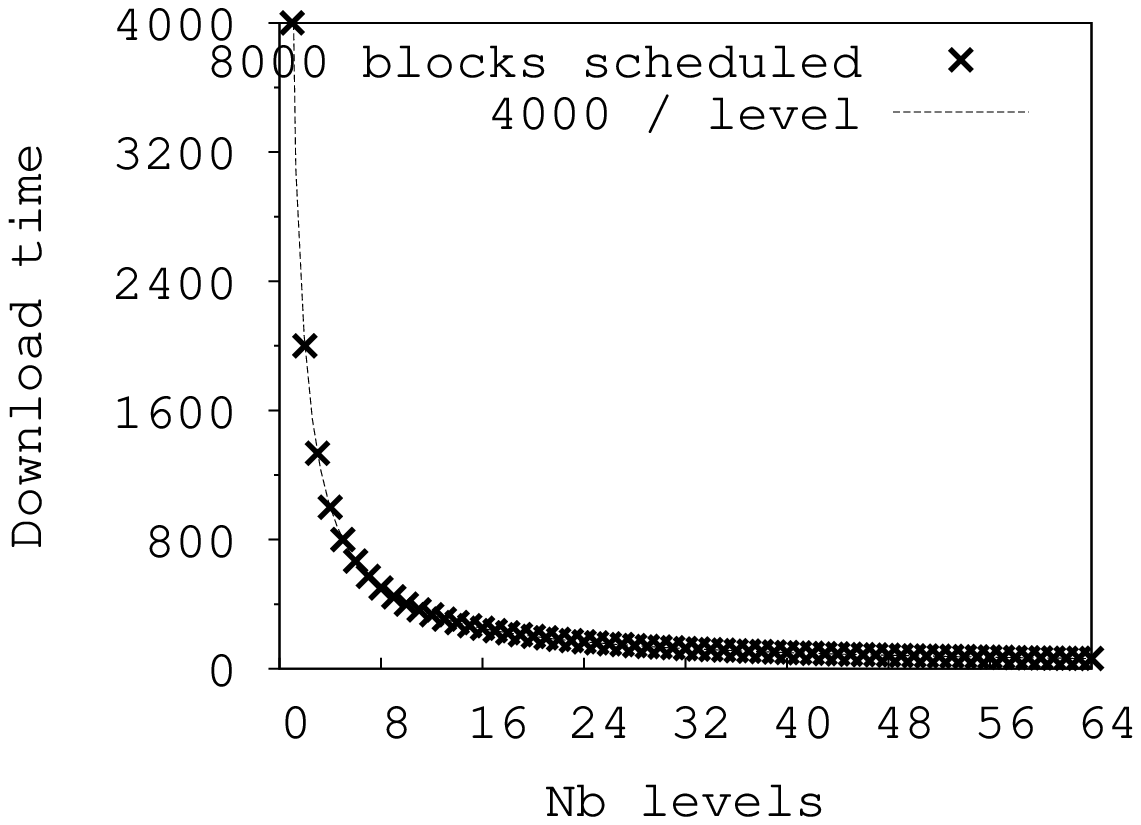}
                \label{fig:theorical_res_sched_fec}
        }
        \caption{Application flow diagram and scheduler performances}
    \end{figure}
    The sequencer can be used by various applications. To this end we implement
    an efficient file transfer with a fine grained rate tuning. Our goal is to
    show the performances and the ease of using a dynamic multicast congestion
    control thanks to the sequencer. This software uses a cyclic data
    transmission scheme named \textit{carousel}
    loop~\cite{peltotalo_performance_2007} and sends data following the flow
    diagram from figure~\ref{fig:soft_data_flow}:
    \begin{itemize}
        \item{
            The file reader splits a file into blocks and give them to the $FEC$
                encoder.
        }
        \item{
            The $FEC$ encoder generates repair symbols from the file blocks also
                named source symbols. The source and repair symbols form
                the $FEC$ symbols.
        }
        \item{
            The application scheduler orders the $FEC$ symbols into hierarchical
                buffers and pass them to the API.
        }
        \item{
            The sequencer orders the hierarchical buffers into packets assigned
                to groups.
        }
        \item{
            The network scheduler is a part of the multicast congestion control.
                It schedules packets for each multicast group.
                $M2C$~\cite{lucas_fair_2009} is used for this evaluation.
        }
    \end{itemize}
    The sequencer coupled with the network scheduler can be reused for other
    applications and provides functionalities similar to $TCP$, except
    reliability which is managed by the $FEC$ encoder and the \textit{carousel}.

    \subsection{Application scheduler}
        The application scheduler, derived from~\cite{birk_multicast_2003},
        orders $B$ blocks into buffers of $N$ hierarchical levels (cf.
                figure~\ref{fig:application_scheduler}), such as the distance
        between two occurrences of a block is maximized and the number of
        redundant blocks received is reduced:
        %\\
        \begin{itemize}
           \item{
               The level $1$ chooses block $b$ for buffer $b$, with $b
                   \in [0...B[$.
           } 
           \item{
               The level $n$ chooses block $b+x$ for buffer $b$. With $n \in
                   [2...N[$ and $x$ the block at the middle of the longest
                   interval between two blocks of all lower levels.  For example
                   in figure~\ref{fig:application_scheduler}, level $3$ chooses
                   block $2$ since it is the block at the middle between block
                   $0$ and $5$ distant of $5$ apart.
               %0 1 2 3 4 5 6 7 8 9
               %
               %0
               %          5
               %    2
               %              7
           } 
        \end{itemize}
        Thus, download time is halved each time the receiver doubles the
        number of levels received (cf.  figure~\ref{fig:theorical_res_sched}).
        Moreover, this property is valid whatever the first buffer $b$ received.
        %\\
        Fine rate granularity is provided by using a block size corresponding to
        the payload of a packet.
        %\\
        Note that the levels used in the application scheduler are independent
        of network scheduler layers.

    \subsection{$FEC$ encoder}
        A $FEC$ code generates $n$ symbols from $k$ source symbols.  The main
        advantage is that only $k + \epsilon$ (with $\epsilon$ small) distinct
        symbols suffice to recover the $k$ source symbols.
        %\\
        \textit{Maximum Distance Separation} ($MDS$) codes ($\epsilon = 0$),
        such as Reed Solomon~\cite{rizzo_effective_1997} codes, have a low
        encoding/decoding rate which limits the
        number of usable symbols.
        %maximal $n$ usable.
        %\\
        Whereas \textit{Low-Density Parity-Check} ($LDPC$) codes ($\epsilon \ne
                0$) release the limitation on the number of usable symbols and
        thus have an high encoding/decoding rate. Our file transfer software
        uses a $LDPC triangle$ code~\cite{roca_design_2003} provided by the
        project "Planete-bcast"~\cite{_planete-bcast:_2006} library.
        %\\
        Besides correcting packet losses, having numerous repair symbols can
        reduce the download time.  The application scheduler analysis (cf.
                table~\ref{table:blocks_sent}) shows that the number of unsent
        blocks before sending the first duplicated block is $< N/2$.  Thus,
        having $n = 2 * k$ symbols improves the application scheduler and ables
        to reduce the downloading time for each new level received (cf.
                figure~\ref{fig:theorical_res_sched_fec}).

\section{File transfer evaluation}
    This section describes the testbed and the evaluation criteria used. Then,
    we analyze the application behavior for two different scenarii:
    $1$ file transfer with $1$ long competing stream and several file transfers
    with background traffic.

    \subsection{Testbed}
        \begin{figure}[t]
            \centering
            \includegraphics[width=0.4\textwidth]{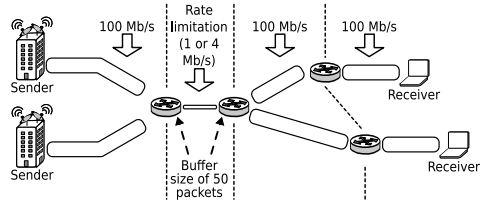}
            \caption{An example of the testbed with 2 senders and 2 receivers}
            \label{fig:testbed}
        \end{figure}
        The file transfer software has been implemented~\cite{_mutual:_2009} and
        evaluated on a testbed (cf.  figure~\ref{fig:testbed}) composed of Linux
        routers running the $XORP$ routing daemon (\textit{PIM-SM},
                \textit{IGMPv3}) coupled with the \textit{Token Bucket Filter}
        ($TBF$) module to set the bottleneck rate ($1$ or $4$ Mb/s) and the
        buffer size (queues of $25$ packets).
        %\\
        As multicast congestion control has a slow rate adaptation, it is best
        suited to manage long lived streams. Then, the size of the files
        transferred must not be too small and we have chosen three files whose
        sizes are:
        \begin{itemize}
            \item{
                $4$ MBytes for a transfer lasting at least respectively $32$ and
                    $8$ seconds at $1$ and $4$ Mb/s.  This does not able $M2C$
                    to converge to the fair rate and shows the behavior of the
                    file transfer with important reception rate variations.
            }
            \item{
                $48$ MBytes for a transfer lasting at least respectively $384$
                    and $96$ seconds at $1$ and $4$ Mb/s. This ables $M2C$ to
                    reach the fair rate.
            }
            \item{
                $99$ MBytes for a transfer lasting at least respectively $792$
                    and $198$ seconds at $1$ and $4$ Mb/s.  This file is the
                    maximal sized manageable by our source and receiver $PC$s,
                            due to memory limits.
            }
        \end{itemize}
        The comparison is done between an unicast download client using $TCP$
        \textit{\mbox{new-RENO}} congestion control algorithm with $SACK$
        option, and a multicast file transfer using our
        implementation~\cite{_mutual:_2009}.

     \subsection{Criteria}
        The file transfer performances evaluation criteria are based on:
        %The criteria used to evaluate the performances of the file transfer
        %are based on:
        \begin{itemize}
            \item{
                $time$: The number of seconds ($s$) used to download and recover
                    the file.
            }
            \item{
                $gput$: The goodput corresponding to the application download
                    rate ($Kb/s$).
            }
            \item{
                $tput$: The throughput corresponding to the link download rate
                    ($Kb/s$).
            }
            \item{
                $loss$: The loss rate ($\%$).
            }
            \item{
                $dup$: The duplicated symbols overhead ($\%$) between the total
                    number of $FEC$ symbols received ($received\_symbols$) and
                    the number of symbols needed by the $FEC$ code to recover
                    the file ($k + \epsilon$): $dup := (received\_symbols / (k +
                                \epsilon) - 1) * 100$.  The sequencer behaves
                                                well, if $dup$ is close to $0$.
            }
            \item{
                $sym$: The $FEC$ symbols overhead ($\%$) needed by the $FEC$
                    code. It is the ratio between the number of source
                    symbols ($k$) and the number of symbols needed
                    by the $FEC$ code to recover the file ($k + \epsilon$):
                        $sym := ((k + \epsilon) / k - 1) * 100$.
            }
            \item{
                $head$: The header overhead ($\%$) between
                    the datagram or packet length ($packet\_length$) and the
                    applicative data ($applicative\_data$) contained by the same
                    packet:
                    \\
                    $head := (packet\_length / applicative\_data - 1) * 100$.
                    \\
                    Since $packet\_length$ is $1480$ and that unicast
                    and multicast carry respectively $1460$ and $1448$ bytes of
                    $applicative\_data$, $head$ is a constant overhead:
                    \begin{itemize}
                        \item{
                            $1.4\%$ for unicast streams, or $3.7\%$ with
                                ethernet and IP headers.
                        }
                        \item{
                            $2.2\%$ for multicast streams, or $4.6\%$ with
                                ethernet and IP headers.
                        }
                    \end{itemize}
            }
            \item{
                $net$: The total network overhead ($\%$). It is the ratio
                    between the length of the file ($file\_length$) and the
                    amount of data received at the link level
                    ($link\_nb\_data$):
                    \\
                    $net := (link\_nb\_data / file\_length - 1) * 100$.
                     \\
                     This criterion represents a general estimation of the
                     total overhead generated in the network and include
                     overhead of $dup$, $sym$ and $head$.
            }
            \item{
                $comp$: The computation time to recover the file after the last
                    $FEC$ symbol is received ($\%$). It is the ratio
                    between the $time$ criterion and the time needed by the
                    network to receive the file ($network\_time$):
                    $comp := (time / network\_time - 1) * 100$.  This criterion
                     represents the time needed to restore the file once all the
                     needed symbols are already received.
            }
        \end{itemize}
        In the following results all these criteria are given with a confidence
        interval of $95\%$ with a series of $20$ tests for each experiment
        setup.

    \subsection{One file transfer with one long competing stream.}
        These tests show the file transfer ($ft$) behavior with a long competing
        stream:
        \begin{itemize}
            \item{
                $1$ \textit{$TCP$ file transfer} ($TCP\_ft$) competing with
                    $1$ $TCP$ stream.
            }
            \item{
                $1$ \textit{$M2C$ file transfer} ($M2C\_ft$) competing with
                    $1$ $TCP$ stream.
            }
            \item{
                $1$ \textit{$M2C$ file transfer} ($M2C\_ft$) competing with
                    $1$ $M2C$ stream.
            }
        \end{itemize}
        \begin{figure}
            \centering
            \subfigure[$1$ file transfer versus $1$ long stream]
            {
                \includegraphics[width=12cm]
                    %{img/res/res_tests_1vs1.eps}
                    {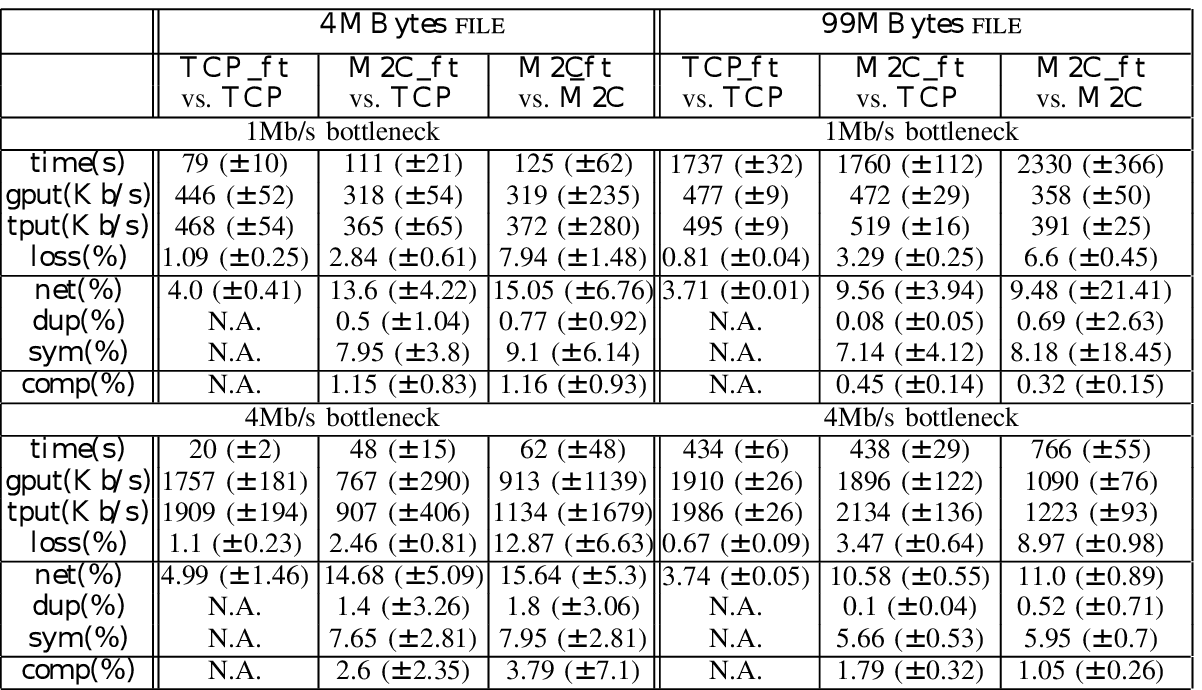}
                \label{tab:m1t1}
            }
            \subfigure[$1$ and $2$ file transfers with background traffic]
            {
                \includegraphics[width=12cm]
                    %{img/res/res_tests_multi.eps}
                    {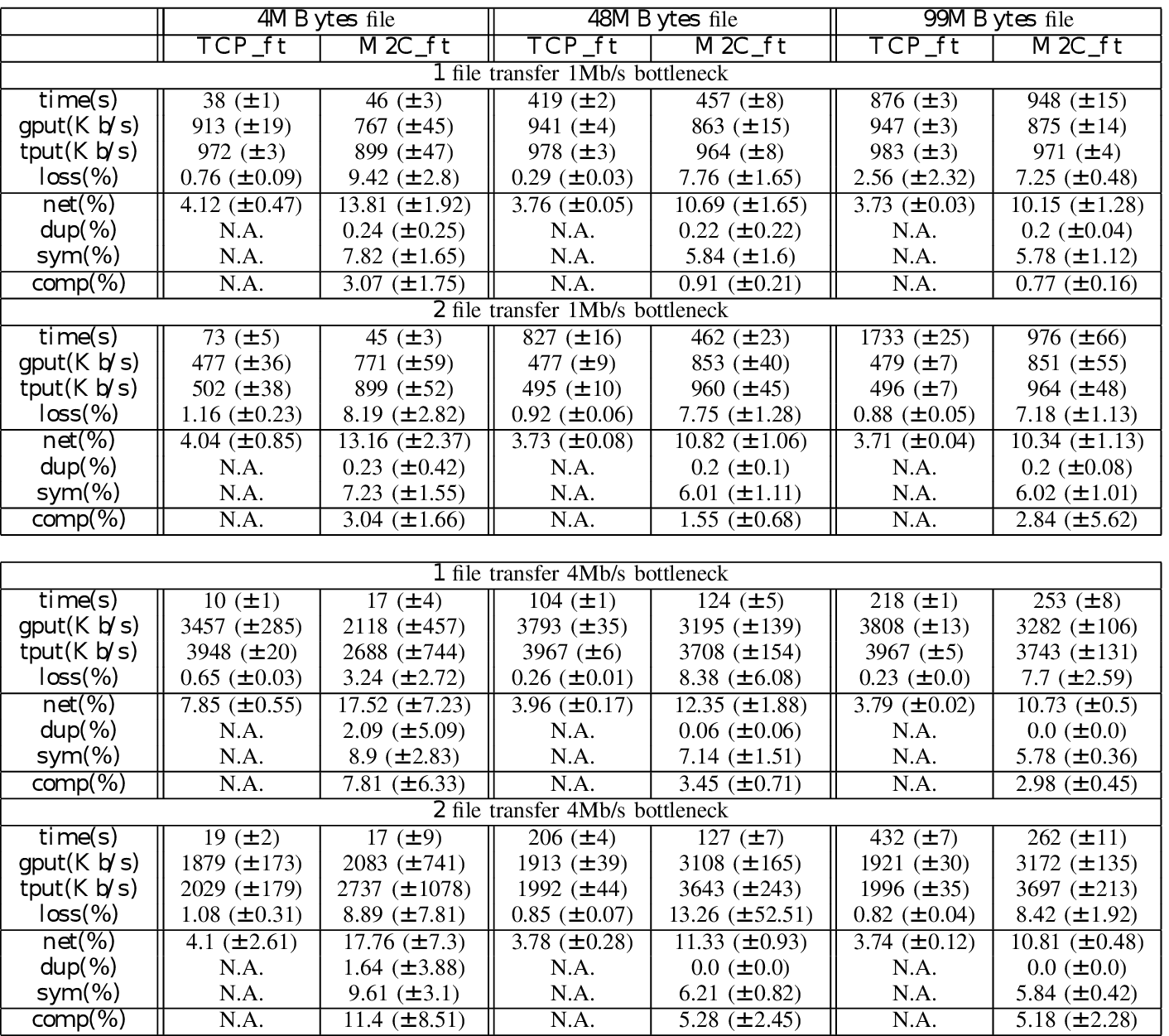}
                \label{tab:multi_1_and_2}
            }
            \subfigure[$5$ file transfers with background traffic for the
                $48MBytes$ file]
            {
        %\begin{table}
        %    %\renewcommand{\arraystretch}{1.3}
        %    \centering
        %    \caption{$5$ file transfers with background traffic for the
        %        $48MBytes$ file.}
            \label{tab:multi_5_48}
            \begin{tabular}{|c|c|c|c|c|}
                \hline
                & $TCP\_ft$
                    & $M2C\_ft$
                    & $TCP\_ft$
                    & $M2C\_ft$
                    \\
                \hline
                & \multicolumn{2}{|c|}{1Mb/s bottleneck}
                    & \multicolumn{2}{|c|}{4Mb/s bottleneck} \\
                \hline
				$time (s)$
					& 2055 ($\pm$30)
					& 495 ($\pm$65)
					& 512 ($\pm$8)
					& 146 ($\pm$28)
					\\
				$gput (Kb/s)$
					& 192 ($\pm$3)
					& 801 ($\pm$97)
					& 770 ($\pm$13)
					& 2733 ($\pm$467)
					\\
				$tput (Kb/s)$
					& 199 ($\pm$3)
					& 921 ($\pm$129)
					& 799 ($\pm$13)
					& 3360 ($\pm$754)
					\\
				$loss (\%)$
					& 5.86 ($\pm$1.85)
					& 9.86 ($\pm$9.5)
					& 4.03 ($\pm$0.11)
					& 8.17 ($\pm$3.36)
					\\
				\hline
				$net (\%)$
					& 3.71 ($\pm$0.06)
					& 11.64 ($\pm$7.27)
					& 3.68 ($\pm$0.26)
					& 11.38 ($\pm$1.41)
					\\
				$dup (\%)$
                    & N.A.
					& 0.51 ($\pm$1.29)
                    & N.A.
					& 0.21 ($\pm$1.08)
					\\
				$sym (\%)$
                    & N.A.
					& 6.49 ($\pm$1.79)
                    & N.A.
					& 6.06 ($\pm$0.93)
					\\
				\hline
				$comp (\%)$
                    & N.A.
					& 3.04 ($\pm$2.67)
                    & N.A.
					& 10.17 ($\pm$7.61)
					\\
                \hline
            \end{tabular}
        %\end{table}
            }
            \caption{Experiments results}
        \end{figure}
        In all these tests, $TCP$ $net$ is less than $5\%$, mainly composed of
        the $3.7\%$ of header overhead.
        %\\
        Table~\ref{tab:m1t1} shows for $99$ MBytes files that $M2C\_ft$ $net$ is
        about $10\%$, mainly composed of at least $5.66\%$ of $sym$ and $4.6\%$
        of header overhead.  Concerning the $4$ MBytes file, $M2C\_ft$ $net$ is
        between $13$ and $16\%$.  The increase of $net$ is linked to a
        growth of additional $FEC$ symbols required ($8\%$). This can be
        explained as the $\epsilon$ extra symbols varies in function of the
        symbols received: all symbols have not the same importance.  Moreover
        with small files, the $tput$ variation is due to $M2C$ slow
        convergence to the fair rate, which confirms that $M2C\_ft$ is not
        suited for too short files.
        %\\
        The small $dup$ values prove the sequencer good behavior.
        %\\
        $TCP$ and $M2C$ streams use an \textit{Additive Increase and
            Multiplicative Decrease} ($AIMD$)~\cite{yang_general_2000} mechanism
            producing cyclic losses.  But, $M2C$ losses are more bursty, due to
            sudden rate variations when joining a new group.  Regardless of the
            obtained throughput, the file transfer is not sensitive to the loss
            distribution and is quite as efficient with $3\%$ of independent
            losses, as with $10\%$ of bursty losses.  Finally, download time
            differences are due to $M2C$ loose fairness when competing with
            another $M2C$ stream.
        %\\
        The file transfer provides a fine grain rate as its behavior is not
        sensitive to the different bottleneck limitations and adapts itself to
        various $tput$.

    \subsection{File transfer with background traffic and multiple receivers.}
        These tests show the benefits of using $M2C\_ft$ when several receivers
        download the same file in a more realistic environment: $1$ or $2$
        simultaneous downloads compete with background traffic composed of many
        short $TCP$ connections such as those used for web browsing. These
        streams start times follow a Poisson process with an average of 10
        streams per minute. Each stream sends an amount of data defined by
        an exponential distribution, such that $85\%$ of the connections carry
        less than $6$KB.
        %\\
        Table~\ref{tab:multi_1_and_2} shows that for both $TCP\_ft$ and
        $M2C\_ft$, $net$ is similar with $1$ or $2$ receivers.  However,
        $TCP\_ft$ download time almost doubles when $2$ receivers are involved,
        whereas $M2C\_ft$ download time remains stable. Thus, $M2C\_ft$ is able
        to take advantage of the multicast scalability, while unicast streams
        are forced to share the bandwidth with each other. Then, only $2$
        receivers are enough to make $M2C\_ft$ more advantageous than $TCP\_ft$.
        %\\
        To show that the benefit of using $M2C\_ft$ increases with the
        number of receivers of the same file, we have done another series of
        tests with $5$ receivers downloading the same $48MBytes$ file competing
        with background traffic. Results shown in table~\ref{tab:multi_5_48}
        confirm $M2C\_ft$ performances. Indeed, these results point out that
        $M2C\_ft$ is almost independent of the number of receivers, while for
        $TCP$ the download time increases linearly with the number of receivers.

%        - Warmly acknowledge people who have helped you
%        - Be generous to the competition. “In his inspiring paper [Foo98]
%           Foogle shows.... We develop his foundation in the following ways...”
%        - Acknowledge weaknesses in your approach
%\section{Previous work}

%        -    Be brief.
\section{Conclusion}
    This paper presents a new sequencer and $API$ in order to ease the use of
    multicast congestion control with dynamic layering. Indeed, despite
    promising results in terms of fairness, using efficiently these dynamic
    groups is a challenging task for application programmers and no solution has
    been proposed yet.
    %\\
    The sequencer maps out application data to dynamic groups in an optimal way.
    Multiple applications such as file transfer or video streaming, can use this
    sequencer, thanks to a simple $API$ which takes as parameter a
    hierarchically encoded buffer, without knowing the hierarchy application
    scheme.  Thereby, receivers get all the most important data they can afford,
    since they join groups from bottom-up due to the cumulative hierarchy used
    by the multicast congestion control. 
    %\\
    To evaluate our proposition, we designed a file transfer using
    this sequencer, a multicast congestion control, an application scheduler and
    a $FEC$ coder.
    %\\
    The evaluation on a testbed shows the optimal behavior of the sequencer and
    the file transfer efficiency.  Indeed, it only produces a slightly larger
    overhead than $TCP$ for a single download.  Furthermore, the multicast file
    transfer is highly scalable, almost independent of the number of receivers,
    and is more advantageous than $TCP$ starting at only $2$ receivers.
    %\\
    Finally, this file transfer receives a fine rate granularity, which
    is totally independent of the group hierarchy used by the multicast
    congestion control. Moreover, each receiver can start downloading at any
    time independently of the source or other receivers.
    %\\
    Currently, the transferred file size is limited by the $PC$ memory.
    Future work will focus on the use of an interleaver to release this
    limitation.
    %\\
    Lastly, the sequencer $API$ is usable by various applications and is already
    integrated in a video streaming software using a
    $PVH$~\cite{mccanne_low-complexity_1997} codec with dynamic source channels.

\bibliographystyle{splncs}
\bibliography{lucas_pansiot_grad_hilt_scheduler}

\end{document}